\documentclass[11pt,a4paper,twoside,twocolumn]{article}

\usepackage{graphicx}
\usepackage{mathpaper}
\usepackage{comment}
\usepackage{color}
\usepackage{url}

\usepackage{a4wide}

\def\papertitle{Live music programming in Haskell}
\def\paperauthorA{Henning Thielemann}

\author{\paperauthorA}

\usepackage{fancyhdr}
\usepackage{hyperref}

\newcommand\term[1]{\hyperref[#1]{\textit{#1}}}
\newcommand\altterm[2]{\hyperref[#1]{\textit{#2}}}
\newcommand\extended[1]{#1}
\newcommand\keywords[1]{\noindent\textbf{Keywords:} #1}

\begin{document}

\newcommand\secpageref[1]{page~\pageref{sec:#1}}     

\newcommand\figcaption[2]{\caption{\textit{#2}}\figlabel{#1}}
\newcommand\tabcaption[2]{\caption{\textit{#2}}\tablabel{#1}}

\newcommand\todo[1]{}

\newcommand\code[1]{\texttt{#1}}

\newenvironment{contentsmall}{\small}

\title{\papertitle}

\sloppy

\maketitle

\begin{abstract}
\begin{contentsmall}
We aim for composing algorithmic music
in an interactive way with multiple participants.
To this end we have developed an interpreter for a sub-language
of the non-strict functional programming language Haskell
that allows the modification of a program during its execution.
Our system can be used both for musical live-coding
and for demonstration and education of functional programming.
\end{contentsmall}
\end{abstract}


\section{Introduction}

It is our goal to compose music by algorithms.
We do not want to represent music as a sequence of somehow unrelated notes
as it is done on a note sheet.
Instead we want to describe musical structure.
For example, we do not want to explicitly list the notes of an accompaniment
but instead we want to express the accompaniment
by a general pattern and a sequence of harmonies.
A composer who wants to draw a sequence of arbitrary notes
might serve as another example.
The composer does not want to generate the random melody note by note
but instead he wants to express the idea of randomness.
Following such a general phrase like ``randomness''
the interpreter would be free to play a different
but still random sequence of notes.

The programmer shall be free to choose the degree of structuring.
For instance, it should be possible
to compose a melody manually,
accompany it using a note pattern
following a sequence of user defined harmonies
and complete it with a fully automatic rhythm.

With a lot of abstraction from the actual music
it becomes more difficult
to predict the effect of the programming on the musical result.
If you are composing music that is not strictly
structured by bars and voices
then it becomes more difficult
to listen to a certain time interval or a selection of voices
for testing purposes.
Also, the classical edit-compile-run loop hinders creative experiments.
Even if the music program can be compiled and restarted quickly,
you must terminate the running program and thus the playing music
and you must start the music from the beginning.
Especially if you play together with other musicians
this is unacceptable.

In our approach to music programming
we use a purely functional \altterm{non-strictness}{non-strict}%
\footnote{All terms set in italics are explained
in the glossary on \secpageref{glossary}.
In the PDF they are also hyperlinks.}
programming language
\cite{hughes1984fpmatter},
that is almost a subset of Haskell~98 \cite{peyton-jones1998haskell}.
Our contributions to live music coding are
concepts and a running system
offering the following:
\begin{itemize}
\item algorithmic music composition
where the program can be altered while the music is playing
(\secref{live-coding}),
\item simultaneous contributions of multiple programmers
to one song led by a conductor
(\secref{multiuser}).
\end{itemize}



\section{Functional live programming}

\subsection{Live coding}
\seclabel{live-coding}

We want to generate music as a list of MIDI events
\cite{mma1996midi},
that is events like
``key pressed'', ``key released'',
``switched instrument'', ``knob turned''
and wait instructions.
A tone with pitch C-5, a duration of 100 milliseconds
and an average force
shall be written as:
\begin{verbatim}
main =
   [ Event (On c5 normalVelocity)
   , Wait 100
   , Event (Off c5 normalVelocity)
   ] ;

c5 = 60 ;
normalVelocity = 64 ;
.
\end{verbatim}
Using the list concatenation ``\verb|++|''
we can already express a simple melody.
\begin{verbatim}
main =
   note qn c ++ note qn d ++
   note qn e ++ note qn f ++
   note hn g ++ note hn g ;

note duration pitch =
   [ Event (On pitch normalVelocity)
   , Wait duration
   , Event (Off pitch normalVelocity)
   ] ;

qn = 200 ;  -- quarter note
hn = 2*qn ; -- half note

c = 60 ;
d = 62 ;
e = 64 ;
f = 65 ;
g = 67 ;
normalVelocity = 64 ;
\end{verbatim}
We can repeat this melody infinitely
by starting it again when we reach the end of the melody.
\begin{verbatim}
main =
   note qn c ++ note qn d ++
   note qn e ++ note qn f ++
   note hn g ++ note hn g ++ main ;
\end{verbatim}
Please note, that this is not a plain recursion,
but a so called \term{co-recursion}.
If we define the list \verb|main| this way
it is infinitely long
but if we expand function applications only when necessary
then we can evaluate it element by element.
Thanks to this evaluation strategy
(in a sense \term{lazy evaluation} without \term{sharing})
we can describe music as pure list of events.
The music program does not need, and currently cannot,
call any statements for interaction with the real world.
Only the interpreter sends MIDI messages to other devices.

In a traditional interactive interpreter
like the \verb|GHCi|%
\footnote{Glasgow Haskell Compiler in interactive mode}
we would certainly play the music this way:
\begin{verbatim}
Prelude> playMidi main       .
\end{verbatim}
If we would like to modify the melody
we would have to terminate it
and restart the modified melody.
In contrast to this we want to alter the melody
while the original melody remains playing
and we want to smoothly lead over from the old melody to the new one.
In other words:
The current state of the interpreter
consists of the program and the state of the interpretation.
We want to switch the program,
but we want to keep the state of interpretation.
This means that the interpreter state must be stored in a way
such that it stays sensible even after a program switch.

We solve this problem as follows:
The interpreter treats the program as a set of term rewriting rules,
and executing a program means to apply rewrite rules repeatedly
until the start term \verb|main| is expanded far enough
that the root of the operator tree is a terminal symbol
(here a \term{constructor}).
For the musical application the interpreter additionally tests
whether the root operator is a list constructor,
and if it is the constructor for the non-empty list
then it completely expands the leading element
and checks whether it is a MIDI event.
The partially expanded term forms the state of the interpreter.
For instance,
while the next to last note of the loop from above is playing,
that is, after the interpreter has sent its \code{NoteOn} event,
the current interpreter state would look like:
\begin{verbatim}
Wait 200 :
   (Event (Off g normalVelocity) :
      (note hn g ++ main))
.
\end{verbatim}

The interpreter will rewrite the current expression as little as possible,
such that the next MIDI event can be determined.
On the one hand this allows us
to process a formally infinite list like \verb|main|,
and on the other hand
you can still observe the structure of the remaining song.
E.g. the final call to \verb|main| is still part of the current term.
If we now change the definition of \verb|main|
then the modified definition will be used when \verb|main| is expanded next time.
This way we can alter the melody within the loop, for instance to:
\begin{verbatim}
main =
   note qn c ++ note qn d ++
   note qn e ++ note qn f ++
   note qn g ++ note qn e ++
   note hn g ++ main ;
.
\end{verbatim}
But we can also modify it to
\begin{verbatim}
main =
   note qn c ++ note qn d ++
   note qn e ++ note qn f ++
   note hn g ++ note hn g ++ loopA ;
\end{verbatim}
in order to continue the melody with another one called \verb|loopA|
after another repetition of the \verb|main| loop.

We want to summarise
that the meaning of an expression can change during the execution of a program.
That is, we give up a fundamental feature of functional programming,
namely \term{referential transparency}.

\todo{stress that we do not send commands, but change the program}

We could implement the original loop
using the standard list function \verb|cycle|
\begin{verbatim}
main =
  cycle
    ( note qn c ++ note qn d ++
      note qn e ++ note qn f ++
      note hn g ++ note hn g ) ;
\end{verbatim}
and if \verb|cycle| is defined by
\begin{verbatim}
cycle xs = xs ++ cycle xs ;
\end{verbatim}
then this would be eventually expanded to
\begin{verbatim}
( note qn c ++ note qn d ++
  note qn e ++ note qn f ++
  note hn g ++ note hn g )
++
cycle
   ( note qn c ++ note qn d ++
     note qn e ++ note qn f ++
     note hn g ++ note hn g ) ;
.
\end{verbatim}
Using this definition we could leave the loop only
by changing the definition of \verb|cycle|.
But such a change would affect \emph{all} calls of \verb|cycle|
in the current term.
Further, in a rigorous module system without import cycles
it would be impossible to access functions of the main module
from within the standard module \verb|List|
that defines the \verb|cycle| function.
But this would be necessary in order to not only leave the \verb|cycle| loop
but to continue the program in the main module.

From this example we learn
that a manually programmed loop
in the form of \verb|main = ... ++ main|
has advantages over a loop function from the standard library,
because the manual loop provides a position
where we can insert new code later.

Additionally to the serial composition of musical events
we need the parallel composition
for the simultaneous playback of melodies, rhythms and so on.
At the level of MIDI commands this means
that the commands of two lists must be interleaved in the proper way.
For details we refer the reader to the implementation of ``\verb|=:=|''.

\subsubsection{User interface}

The graphical user interface is displayed in \figref{screenshot}.
In the upper left part the user enters the program code.
Using a keyboard short-cut
he can check the program code and
transfer it to the buffer of the interpreter.
The executed program is shown in the upper right part.
In this part the interpreter highlights
the function calls that had to be expanded
in order to rewrite the previous interpreter term into the current one.
This allows the user to trace the melody visually.
The current term of the interpreter is presented
in the lower part of the window.
The texts in the figure are essentially the ones
from our introductory example.
\begin{figure}[htb]
  \begin{center}
    \includegraphics[width=\hsize]{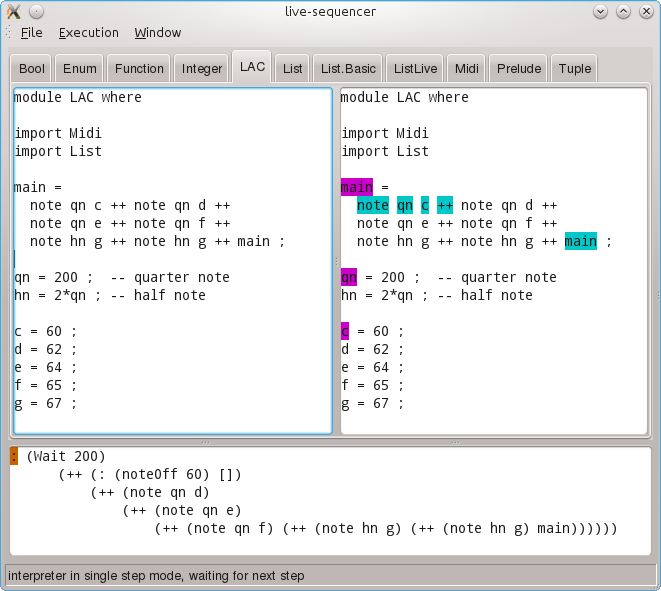}
    \figcaption{screenshot}{The running interpreter}
  \end{center}
\end{figure}

Our system can be run in three modes:
``real time'', ``slow motion'' and ``single step''.
The real-time mode plays the music as required by the note durations.
In contrast to that the other two modes
ignore the wait instructions
and insert a pause after every element of the MIDI event list.
These two modes are intended for studies and debugging.
You may also use them in education
if you want to explain how an interpreter
of a non-strict functional language works in principle.

We implemented the interpreter in Haskell
using the Glasgow Haskell Compiler GHC \cite{ghc2012},
and we employ WxWidgets \cite{smart2011wxwidgets}
for the graphical user interface.
Our interpreted language supports
pattern matching, a set of predefined infix operators,
higher order functions, and partial function application.
For the sake of a simple implementation
we deviate from Haskell 98 in various respects:
Our language is dynamically and weakly typed:
It knows ``integer'', ``text'' and ``\term{constructor}''.
The parser does not pay attention to layout
thus you have to terminate every declaration with a semicolon.
Several other syntactic features of Haskell 98 are neglected,
including list comprehensions, operator sections, do notation,
``let'' and ``case'' notation, and custom infix operators.
I/O operations are not supported as well.

\subsection{Distributed coding}
\seclabel{multiuser}

Our system should allow
the audience to contribute to a performance
or the students to contribute to a lecture
by editing program code.
The typical setup is
that the speaker projects the graphical interface of the sequencer
at the wall,
the audience can listen to music through loud speakers,
and the participants can access the computer of the performer
via their browsers and wireless network.

Our functional language provides a simple module system.
This helps the performer to divide a song into sections or tracks
and to put every part in a dedicated module.
Then he can assign a module to each participant.
This is still not a function of the program,
but must be negotiated through other means.
For instance the conductor might point to people in the audience.
Additionally the performer can insert a marker comment
that starts the range of text that participants can edit.
The leading non-editable region will usually contain the module name,
the list of exported identifiers,
the list of import statements,
and basic definitions.
This way the performer can enforce an interface for every module.

A participant can load a module into his web browser.
The participant sees an HTML page
showing the non-editable header part as plain text
and the editable region as an editable text field.
(cf.~\figref{browser})
After editing the lower part of module
he can submit the modified content to the server.
The server replaces the text below the marker comment
with the submitted text.
Subsequently the new module content is checked syntactically
and on success it is loaded into the interpreter buffer.
In case of syntax errors in the new code
the submitted code remains in the editor field.
The performer can inspect it there and can make suggestions for fixes.
\begin{figure}[htb]
  \begin{center}
    \includegraphics[width=\hsize]{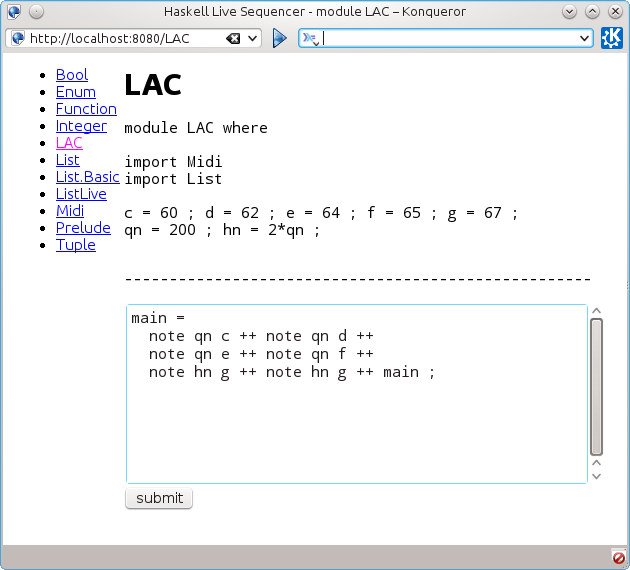}
    \figcaption{browser}{Accessing a module via HTTP}
  \end{center}
\end{figure}

Generally it will not be possible
to start composition with many people from scratch.
However, the performer can prepare a session
by defining a set of modules
and filling them with basic definitions.
For instance he can provide a function
that converts a list of zeros and ones into a rhythm,
or a list of integers into a chord pattern or a bass line.
By providing a meter and a sequence of harmonies he can assert
that the parts contributed by the participants fit together loosely.
In this application the performer no longer plays the role of the composer
but the role of a conductor.




\subsection{Timing}

For a good listening experience
we need precise timing for sending MIDI messages.
A naive approach would be to send the messages as we compute them.
I.e. in every step we would determine the next element in the list of MIDI events.
If it is a wait instruction then we would wait for the desired duration
and if it is a MIDI event then we would send it immediately.
However this leads to audible inaccuracies
due to processor load caused
by term rewriting, garbage collection and GUI updates.

We are using the ALSA sequencer interface for sending MIDI messages.
It allows us to send a MIDI message with a precise but future time stamp.
However we still want that the music immediately starts
if we start the interpreter
and that the music immediately stops if we stop it
and that we can also continue a paused song immediately.
We achieve all these constraints the following way:
We define a latency, say $d$ milliseconds.
The interpreter will always compute as many events in advance
until the computed time stamps are $d$ milliseconds ahead of current time.
This means that the interpreter will compute a lot without waiting
when it is started.
It will not immediately send a MIDI message because
it needs to compute it first.
This introduces a delay, sometimes audible, but we cannot do it faster.
When the user pauses the interpreter,
we halt the timer of our outgoing ALSA queue.
This means that the delivery of messages is immediately stopped,
but there are still messages for the next $d$ milliseconds in the queue.
If the interpreter is continued these messages will be send
at their scheduled time stamps.
If the interpreter is stopped
we simply increase the time of the ALSA queue by $d$ milliseconds
in order to force ALSA to send all remaining messages.

\section{Related work}

Algorithmic composition has a long tradition.
The musical dice games by Mozart
and the Illiac Suite \cite{hiller1959illiacsuite}
might serve as two popular examples here.
Further on, the Haskore project (now Euterpea) \cite{hudak1996haskore}
provides a method for music programming in Haskell.
It also lets you control synthesisers via MIDI
and it supports the generation of audio files via CSound, SuperCollider
or pure Haskell audio signal synthesis.
Like our approach, Haskore relies upon lazy evaluation
which allows for an elegant definition
of formally big or even infinite songs
while its interpretation actually consumes only a small amount of memory.
However, the creative composition process is made more difficult
by the fact that you can listen to a change to a song
only after terminating the old song and starting the new one.
In a sense, our system is an interactive variation of Haskore.

So-called functional reactive programming
is a very popular approach for programming 
of animations, robot controls, graphical user interfaces
and MIDI processing in Haskell
\cite{elliott1997fran}.
Functional reactive programming mimics
working with a time ordered infinite list of events.
But working with actual lazy lists leads
to fundamental problems in real-time processing,
e.g. if a stream of events is divided first but merged again later.
This is a problem that is solved by functional reactive programming libraries.
The advantage of functional reactive MIDI processing
compared to our approach is
that it allows the processing of event input in realtime.
The disadvantage is that usually
you cannot alter a functional reactive program during execution.

Erlang is another functional (but not purely functional) programming language
that accepts changes to a program while the program is running
\cite{armstrong1997erlang}.
Erlang applies eager evaluation.
That is, in Erlang you could not describe a sequence of MIDI commands
by a lazy list of constructors.
Instead you would need iterators or similar tools.
You can insert new program code into a running Erlang programming
in two ways:
Either the running program runs functions (e.g. lambda expressions)
that it receives via messages
or you replace an Erlang module by a new one.
If you upload a new Erlang module
then the old version is kept in the interpreter
in order to continue the running program.
Only calls from outside the module
jump into the code of the new module,
but by qualification you can also simulate an external call
from within the replaced module.
That is, like in our approach, you need dedicated points
(external calls or calls of functions received via messages)
where you can later insert new code.

Summarised, our approach for changing running programs
is very similar to ``Hot Code loading'' in Erlang.
However, the non-strict evaluation of our interpreter implies
that considerable parts of the program are contained in the current term.
These are not affected immediately by a change to the program.
This way we do not need to hold two versions of a module in memory
for a smooth transition from old to new program code.
In a sense, Erlang's external calls play the role of our top-level functions.

\todo{Java Code hot code swapping in Eclipse Debugger?}

Musical live coding,
i.e. the programming of a music generating program,
while the music is playing,
was in the beginning restricted to special purpose languages like
SuperCollider/SCLang \cite{mccartney1996supercollider}
and
ChucK \cite{wang2004chuck}
and their implementations.
With respect to program control
these languages adhere to the imperative programming paradigm
and with respect to the type system
they are object oriented languages.
The main idea in these languages for creating musical patterns
is constructing and altering objects at runtime,
where the objects are responsible for sending commands
to a server for music generation.

Also in our approach the sound generation
runs parallelly to the interpreter
and it is controlled by (MIDI) commands.
However, in our approach
we do not program how to change some runtime objects
but instead we modify the program directly.


In the meantime also Haskell libraries for live coding are available,
like Tidal (\cite{mclean2010tidal}) and
Conductive (\cite{bell2011conductive}).
They achieve interactivity by running commands from
the interactive Haskell interpreter GHCi.
They are similar to SCLang and ChucK in the sense
that they maintain and manipulate (Haskell) objects at runtime,
that in turn control SuperCollider or other software processors.




\section{Conclusions and future work}

Our presented technique demonstrates a new method
for musical live coding.
Maybe it can also be transferred to the maintenance
of other long-running functional programs.
However, we have shown that the user of the live-sequencer
must prepare certain points for later code insertion.
Additionally our system must be reluctant
with automatic optimisations of programs
since an optimisation could remove such an insertion point.
If you modify a running program
then functions are no longer
\altterm{referential transparency}{referentially transparent};
that is, we give up a fundamental feature of functional programming.

\paragraph{Type system}
A static type checker would considerably reduce the danger
that a running program must be aborted
due to an ill-typed or inconsistent change to the program.
The type checker would not only have to test
whether the complete program is type correct after a module update.
Additionally it has to test whether the current interpreter term
is still type correct with respect to the modified program.

A type checker is even more important for distributed composition.
The conductor of a multi-user programming session
could declare type signatures
in the non-editable part of a module
and let the participants implement the corresponding functions.
The type checker would assert
that participants could only send modifications
that fit the rest of the song.

\paragraph{Evaluation strategy}
Currently our interpreter is very simple.
The state of the interpreter is a term that is a pure tree.
This representation does not allow for \term{sharing}.
E.g. if \verb|f| is defined by \verb|f x = x:x:[]|
then the call \verb|f (2+3)| will be expanded to \verb|(2+3) : (2+3) : []|.
However, when the first list element is evaluated to \verb|5|,
the second element will not be evaluated.
I.e. we obtain \verb|5 : (2+3) : []| and not \verb|5 : 5 : []|.
Since the term is a tree and not a general graph
we do not need a custom garbage collector.
Instead we can rely upon the garbage collector
of the GHC runtime system that runs our interpreter.
If a sub-term is no longer needed
it will be removed from the operator tree
and sooner or later it will be detected and de-allocated
by the GHC garbage collector.

Even a simple co-recursive definition
like that of the sequence of Fibonacci numbers
\begin{verbatim}
main = fix fibs
fibs x = 0 : 1 : zipWith (+) x (tail x)
fix f = f (fix f)
\end{verbatim}
leads to an unbounded growth of term size
with our evaluation strategy.
In the future we want to add more strategies
like the graph reduction using the STG machine
\cite{peyton-jones1992stg}.
This would solve the above and other problems.
The operator tree of the current term
would be replaced by an operator graph.
The application of function definitions
and thus the possibility of live modifications of a definition
would remain.
However, in our application there is the danger
that program modification may have different effects
depending on the evaluation strategy.
On the one hand, the sharing of variable values
at different places in the current term
would limit the memory consumption in the Fibonacci sequence defined above,
on the other hand it could make it impossible
to respect a modification of the called function.

Our single step mode would allow
the demonstration and comparison of evaluation strategies in education.

Currently we do not know,
whether and how we could embed our system,
including live program modifications,
into an existing language like Haskell.
This would simplify the study of the interdependence
between program modifications, optimisations and evaluation strategies
and would provide many syntactic and typing features for free.

For this purpose we cannot use an interactive Haskell interpreter like GHCi directly:
\begin{itemize}
\item GHCi does not let us access or even modify a running program
and its internal representation is optimized for execution
and it is not prepared for changes to the running program.
\item GHCi does not allow to observe execution of the program,
and thus we could not highlight active parts in our program view.
\item GHCi does not store the current interpreter state in a human readable way
that we can show in our display of the current term.
\end{itemize}
Nonetheless, we can imagine that it is possible
to write an embedded domain specific language.
That is, we would provide functions that allow to program
Haskell expressions that only generate an intermediate representation
that can then be interpreted by a custom interpreter.


\paragraph{Highlighting}
We have another interesting open problem:
How can we highlight program parts according to the music?
Of course, we would like to highlight the currently played note.
Currently we achieve this by highlighting all symbols
that were reduced since the previous pause.
However if a slow melody
is played parallelly to a fast sequence of controller changes
this means that the notes of the melody are highlighted only for a short time,
namely the time period between controller changes.
Instead we would expect that the highlighting of one part of music
does not interfere with the highlighting of another part of the music.

We can express this property formally:
Let the serial composition operator \verb|++| and
the parallel composition operator \verb|=:=|
be defined both for terms and for highlighting graphics.
Consider the mapping \verb|highl|,
that assigns a term to its visualisation.
Then for every two musical objects \verb|a| and \verb|b| it should hold:
\begin{verbatim}
highl (a ++  b)  =  highl a ++  highl b
highl (a =:= b)  =  highl a =:= highl b
\end{verbatim}
If you highlight all symbols
whose expansion was necessary for generating
a \verb|NoteOn| or \verb|NoteOff| MIDI command,
then we obtain a function \code{highl} with these properties.
However this causes accumulation of highlighted parts.
In
\begin{verbatim}
note qn c ++ note qn d ++
note qn e ++ note qn f
\end{verbatim}
the terms \verb|note qn c| and \verb|note qn d| would still be highlighted
if \verb|note qn e| is played.
The reason is that \verb|note qn c| and \verb|note qn d|
generate finite lists and this is the reason
that \verb|note qn e| can be reached.
That is the expansion of \verb|note qn c| and \verb|note qn d|
is necessary to evaluate \verb|note qn e|.



\paragraph{JACK support}

In the future our system should support JACK in addition to ALSA.
It promises portability and
synchronous control of multiple synthesisers.

\paragraph{Beyond MIDI}

MIDI has several limitations.
For example, it is restricted to 16 channels.
In the current version of our sequencer
the user can add more ALSA sequencer ports
where each port adds 16 virtual MIDI channels.
E.g. the virtual channel 40 addresses the eigth channel of the second port
(zero-based counting).
MIDI through wires is limited to sequential data,
that is, there cannot be simultaneous events.
In contrast to that the ALSA sequencer supports simultaneous events
and our Live sequencer supports that, too.
\extended{
Pitches in MIDI are designed towards 12-tone equal temperament.
You have to use pitch-bending or different interpretation of pitches
in order to relax this bias.
}

Thus the use of MIDI is twofold:
On the one hand it is standard in hardware synthesisers
and it is the only music control protocoll supported by JACK.
On the other hand it has limitations.
The Open Sound Control protocol lifts many of these limitations.
It should also be relatively simple to add OSC support,
but currently it has low priority.

\extended{
We could also think about direct support
of software synthesisers like Csound and SuperCollider.
However we have intentionally chosen to separate the music controller
from the music generator.
This seems to be more modular and keeps the sequencer simple.
Anyway we should think about how to enable Live-sequencer modules
to send e.g. sound algorithms to Csound or SuperCollider.
This way, a Live-sequencer song could carry all information
for reproduction of a song.
}

\section{Acknowledgments}

This project is based on an idea by Johannes Waldmann,
that we developed into a prototype implementation.
I like to thank him, Renick Bell, Alex McLean,
and the anonymous reviewers
for their careful reading and several suggestions for improving this article.

You can get more information on this project
including its development, demonstration videos, and papers
at
\begin{quote}
\url{http://www.haskell.org/haskellwiki/Live-Sequencer} .
\end{quote}

\appendix

\section{Glossary}
\seclabel{glossary}

\paragraph{A constructor}
\label{constructor}
is, mathematically speaking, an injective function
and, operationally speaking, a way to bundle and wrap other values.
E.g. a list may be either empty,
then it is represented by the empty list constructor \verb|[]|,
or it has a leading element,
then it is represented by the constructor \verb|:| for the non-empty list.
For example, we represent a list containing the numbers 1, 2, 3
by \verb|1 : (2 : (3 : []))|,
or more concisely by \verb|1 : 2 : 3 : []|,
since the infix \verb|:| is right-associative.

\paragraph{Co-recursion}
\label{co-recursion}
is a kind of inverted recursion.
Recursion decomposes a big problem into small ones.
E.g. the factorial ``$!$'' of a number
can be defined in terms of the factorial of a smaller number:
\[
n! =
\begin{cases}
1 &: n=0 \\
n \cdot (n-1)! &: n>0
\end{cases}
\]
A recursion always needs a base case,
that is, a smallest or atomic problem
that can be solved without further decomposition.

In contrast to this, co-recursion solves a problem
assuming that it has already solved the problem.
It does not need decomposition and it does not need a base case.
E.g. a co-recursive definition of an infinite list
consisting entirely of zeros is:
\begin{verb}
zeros = 0 : zeros
\end{verb}

\paragraph{Lazy evaluation}
\label{lazy evaluation}
is an evaluation strategy for \altterm{non-strictness}{non-strict semantics}.
An alternative name is ``call-by-need''.
It means that the evaluation of a value is delayed until it is needed.
Additionally it provides \term{sharing} of common results.

\paragraph{Non-strict semantics}
\label{non-strictness}
means that a function may have a defined result
even if it is applied to an undefined value.
It is a purely mathematical property
that is independent from a particular evaluation strategy.

E.g. the logical ``and'' operator \verb|&&| in the C programming language
is non-strict.
In a strict semantics the value of \verb|p && *p|
would be undefined if \code{p} is \code{NULL},
because then \code{*p} would be undefined.
However, \verb|&&| allows the second operand to be undefined
if the first one is \code{false}.

\paragraph{Referential transparency}
\label{referential transparency}
means that function values depend entirely on their explicit inputs.
You may express it formally by:
\[
\forall x, y:\quad x=y \implies f(x)=f(y) \quad.
\]
For mathematical functions this is always true,
e.g. whenever $x=y$ it holds $\sin x = \sin y$.
However for sub-routines in imperative languages this is not true,
e.g. for a function \code{readByte} that reads the next byte from a file,
\code{readByte(fileA)} may differ from \code{readByte(fileB)}
although \code{fileA} = \code{fileB}.

\paragraph{Sharing}
\label{sharing}
means that if you read a variable multiple times
it is still computed only once and then stored for later accesses.

\bibliographystyle{abbrv}
\bibliography{thielemann,audio,haskell,literature}

\end{document}